\documentclass[12pt,a4paper]{article}
\usepackage[latin1]{inputenc}
\usepackage[T1]{fontenc}
\usepackage[english]{babel}
\usepackage{a4,array,latexsym}
\usepackage{graphics}

\newcommand{\bc}{\begin{center}}
\newcommand{\ec}{\end{center}}
\newcommand{\bd}{\begin{displaymath}}
\newcommand{\ed}{\end{displaymath}}
\newcommand{\be}{\begin{equation}}
\newcommand{\ee}{\end{equation}}
\newcommand{\ba}{\begin{array}}
\newcommand{\ea}{\end{array}}
\newcommand{\bea}{\begin{eqnarray}}
\newcommand{\eea}{\end{eqnarray}}
\newcommand{\bt}{\begin{tabular}}
\newcommand{\et}{\end{tabular}}

\newcommand{\bp}{\begin{picture}}
\newcommand{\ep}{\end{picture}}
\newcommand{\bfi}{\begin{figure}}
\newcommand{\efi}{\end{figure}}
\begin{document}


\title{{\huge \bf PREdicted the Higgs Mass
 }}

\author{ 
H.B.~Nielsen 
\footnote{\large\, hbech@nbi.dk} 
\itshape{
The Niels Bohr Institute, Copenhagen, 
Denmark}}


\date{}

\maketitle
This article is one of the contributions 
of Holger Bech Nielsen to the conference
of ``What Comes Beyond the Standard 
Models'' in Bled 2012.

\begin{abstract}
We like to investigate the idea of 
taking  as non-accidental a remarkably
good agreement of our (C.D. Froggatt 
and myself, and also with Yasutaka
Takanishi\cite{MetaMPP}) prediction\cite{Higgsprediction} of the Higgs mass.
 Our modernized most simple
 ``multiple
point principle'' prediction $129.4 \pm 
2$ GeV\cite{Degrassi} versus the recently\cite{CMSHiggs}\cite{ATLASHiggs} observed
$126 \pm 1$ GeV agrees well.

The PREdicted  
Higgs mass is
essentially the smallest 
value, that would 
not make  
our present vacuum  
unstable.

There are two slightly different 
versions in as far as we can {\em either 
} use 
absolute stability by the alternative 
vacuum being required to have higher 
energy density than the present one {\em 
or}  just metastability requiring that 
our vacuum should not have decayed in the 
early time just after big bang (or later
if that should be easier).  
This is of course provided 
we suppose that the 
Standard Model would function almost all 
the way up to the scale relevant for the
alternative vacuum (which as we shall see
is close to the Planck energy scale for 
the Higgs field expectation value).
This is very close to the
suggestion of Akhani-Hamed et al., which 
though rather than the past stability 
discuss the future\cite{Nima}.
Also Shaposhnikov et al have predicted
the Higgs mass.
If this coincidence shows up with higher 
accuracy,
so that we indeed
had to take it as not being just an 
accident, we would 
have to conclude that there should in 
fact not be 
any {\em disturbing} ``new physics'' all 
the way up of such 
a strength, that it 
could disturb the coincidence.
That is to say
the effective potential 
for the Higgs field should not 
 be changed to a degree bigger than 
the accuracy 
with which the coincidense is 
established. In fact some calculations 
\cite{Nima} suggest that exactly 
the potentially by LEP\cite{LEPHiggs}
 found Higgs 
with mass 115 GeV would fit very 
accurately as the smallest Higgs mass 
avoiding our present vacuum blowing into 
a new one.
It would be 
also a confirmation of what we 
called ``multiple 
point principle''. It is finally pointed 
out that 
precisely by using this principle 
C. D. Froggatt 
and I have already found an idea for 
explaining dark 
matter\cite{boundfirst, bound} inside a pure Standard Model. Thus 
a picture
with all new physics postponed to being 
only close 
to the Planck energy scale, except for 
some see-saw 
neutrinos with masses in an intermediate 
range, e.g. 
$10^{12} \ GeV$ is supported. If the 
see-saw neutrinoes 
couple 
reasonable weakly 
they will though not modify essentially 
the running of 
the Higgs self coupling $\lambda$ and the 
thereby 
associated Higgs effective potential. If 
the inflaton 
field could indeed be the Higgs field 
itself as has been 
discussed, although not so successfully 
though, there 
would be no further need for new physics 
to an 
essential degree
up to almost the Planck scale!

The principle we used to our PREdiction 
were the requirement of degenerate vacua 
which we called ``multiple point 
principle''\cite{MPP} and I shall seek 
to deliver 
some arguments that this MPP is ``nice''
to assume, i.e. it is likely to be true. 
\end{abstract}

\newpage
\thispagestyle{empty}
\section{Introduction}

The major starting point of the present 
contribution to the Bled conference is, 
that the recent observation of the Higgs 
mass as observed at CMS\cite{CMSHiggs}  
and ATLAS\cite{ATLASHiggs} at LHC 
lies exceeding tight to the stability
boarder  
of the vacuum in which we live.
That the Higgs mass should indeed lie 
{\em either} at the border at which 
the energy 
density of our present vacuum and the 
alternative vacuum with Higgs field vacuum
expectation value close to the Planck 
energy are the same\cite{Higgsprediction}
 {\em or} where  our present vacuum 
only barely survived the era shortly after
Big Bang (or the early time) without 
transforming itself into the alternative 
vacuum\cite{MetaMPP} were already 
publiched as or PREdicted years ago.
This prediction(s) were based on the 
assumption of ``Multiple Point Principle''
(=MPP) proposed by D. Bennett and myself
\cite{MPP}.

The point 
is indeed that for an even lower Higgs 
mass than these $129.4$ GeV\cite{Degrassi}
 the 
extrapolation of the effective potential 
$V_{eff}(\phi_h)$
using purely Standard Model would lead 
 to 
a negative effective potential.
Various other discussion of the 
Higgs and Fermion masses from cosmological
restriction are found in \cite{Cosmo, r18,
r19,r20,r21,r22,r23,r24,r25,r26,r27,r28,
r29,r30,r31,r32,r33,r34,r35,r36,r37, 
r38}. 

Also Shaposnikov et al 
\cite{Shaposnikov} have predited the 
Higgs mass using it as inflaton, 
It may be explainable that 
our predictions and Shaposnikovs et al.'s are close by both having 
Standard Model high up in energy.

In Michael Scherer's dissertation\cite{Scherer} 
one finds a Higgs mass $\sqrt{6.845}v$ (where $v$ is
the Higgs field vacuum expectation value).

 Sylwester Kornowski 
rather postdict \cite{Kornowski} the Higgs mass. 

A bit depending on the temperature 
in the cosmological eras to be passed 
a sufficiently low Higgs mass would lead 
to the next minimum in the effective 
potential being so low that the vacuum 
we live in would have decayed.    
\begin{center}
\includegraphics{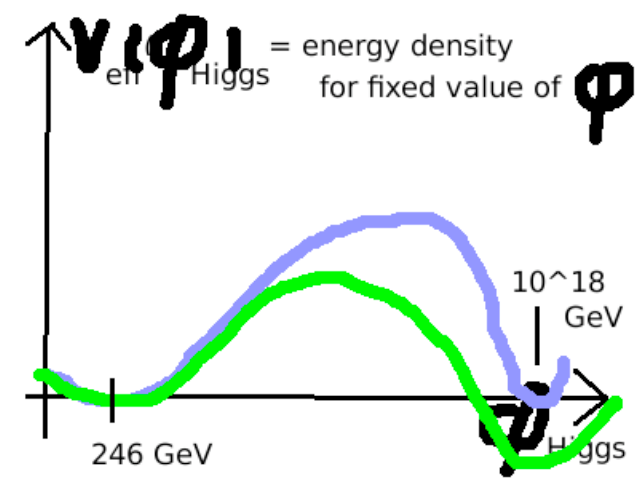} 
\end{center}

For higher mass than about the 129.4 GeV 
the effective potential  will be 
positive all over.

The main point of the present article is 
to say, that, 
if this measurement of the Higgs mass 
should in the 
longer run turn out to be indeed with 
high accuracy 
{\em just} the lowest value needed for 
the metastability or stability 
of the present vacuum, that would be a
remarkable 
coincidense. Thus we should take it 
serious in the 
sense that we should say it is not an 
accident, but 
due to some physical effect causing this 
borderline 
value. But if that is so, then of course 
it also
requires that the competing vacuum in fact
at a value  
of a Higgs field near $10^{18}$ GeV would 
have to 
indeed exist. Otherwise how could it 
have any effect? But 
then it would mean that the Standard 
Model should only 
be tinily modified up to this Higgs 
field value / energy
scale relevant for the alternative 
vacuum. So if indeed such Higgs mass 
value should be 
accurately enough measured to have such a 
specially 
significant 
value, then we must also believe the 
Standard Model
to work so high in energy as to make the 
specially 
significant value 
indeed be significant. 
We shall see in section\ref{SMatw} 
that we {\em can indeed imagine that 
the Standard Model could be valid 
all the way up to almost the Planck scale
in energy scale (or equivalently say 
field strength scale) except that we
still must accept the existence of 
some right handed neutrinoes - see saw 
neutrinoes - to allow for neutrino-
oscillations and baryonnumber excess 
needed for getting enough baryons
minus anti baryons}.

Before going on to this attempt to rescue
a picture of physical laws with Standard 
Model all the way up in energy to about 
one or two orders of magnitude under the 
Planck scale except for some ``see-saw''
or right handed neutrinoes we shall 
in the next section \ref{support} put
forward some propaganda-like arguments
for that the ``multiple point principle''
telling that there are degenenerate vacua 
is not quite as arbitrary and unjustified
as it may seem at first.

To put the ``Multiple Point Principle''
in the right perspective we must remind 
the reader that modern high energy 
physics is plagued or mystified by a 
series of {\em fine tuning problems}:
That is to say there are some parameters 
- meaning coupling constants or other 
coefficients in the Lagrange density -
in the laws of nature or say in the 
Lagrangean density which take on very 
simple / special  value-combinations. 
Or rather the parameters
in the Lagrangian are usually corrected 
by divergent corrections to give the 
``renormalized parameters'' which are 
the measurable ones, and it is these 
{\em renormalized parameters that take 
on the values that are so simple / special
 that they require an explanation.}

The most dramatic example is the 
cosmological constant (= the dark energy
= the energy density of vacuum)
which now although not exactly zero 
is enormously small compared to the 
order of magnitude of the various 
contributions one can imagine to 
contribute to this vacuum energy density
such as a contrbution from the Higgs field,
or simply divergences with a cut off put 
at the Planck scale.

Also famous is the problem of the Higgs 
mass and weak interaction energy scale       being enormously small compared to 
Planck scale or GUT scale, if such one 
should exist(actually the present article
leads to the weak prediction that there is 
no GUT).

Let me stress, that it is one of the 
major points of the present article
to suggest, that rather than attempting 
 -as most high energy physicists - to 
somehow
dream up symmetries and circumvent 
the fine tuning problems without having 
to postulate that the couplings are truly
finetuned, one should rather 
indeed finetune. The success of this 
program
of avoiding fine tuned couplings seems 
not to have been though exceedingly 
successfull. When I say this, I am just 
refering to that e.g. the hierarchy 
problem, which can be considered a part
of the weak to Planck scale ratio problem
seems to require new physics compared to 
the Standard Model with supersymmetric 
partner particles so far not seen. For  
the problem of the cosmological constant
the most promissing solution seems to be 
antropic principle, which really is more
like a finetuning postulate. One must have
lots of universes with different couplings
and then only those which have such 
combinations that we humans can live 
there are in practice realized. Well, in 
principles there are also some without 
humans, but  the antropic principle 
functions in practice like a machine 
or principle settling the couplings to be 
just so as to allow the humans. So one 
might say that in this way the antropic 
principle functions as a finetuner 
mechanism rather than avoiding finetuning.

The point of view of the present article
is now:

{\bf Give up to avoid finetuning and 
then rather look for a fine tuning 
principle telling how the coupling 
constants or other parameters are 
fine tuned; a theory of finetuning,
not avoiding finetuning!}

That is to say, we take here the 
point of view that, since it is too 
hard to avoid finetuning, we shall instead
look for some simple rules about what 
values the coupling constant or 
similar parameters take on. In principle
we suggest that physicists shall make 
a series of attempt-models each delivering
its set of coupling constants or some 
restrictions just on the system of 
couplings and then simply look 
phenomenologically on the sets of such
proposed systems of coupling constant 
restrictions to find out which 
restriction-system is the simplest and 
most 
beautyfull compared to how strong 
restrictions it can provide. Then
we should take the ``right'' system
of coupling restrictions  to be the 
in this sense  nicest system of 
couplings 
with relatively strong restriction. 
But then if one were successful to find 
such a system of coupling restrictions, 
then one would have to believe, 
that there indeed exists a law of nature
providing such a coupling constant 
restriction system. 

The ``multiple point principle'' is 
a concrete proposal for such a rather
promissing system for restriction of coupling 
constants. 
That is to say the ``multiple
point principle'' is a proposal for what
a law for specifying coupling constants
and similar parameters could be. 
From the  multiple point principle we
obtain restrictions between parameters,
which according to the already stressed
example of the Higgs mass seem 
to have at least one example it which it 
is confirmed 
to be right experimentally.

(That we truly made a PREdiction can be seen from that I were even painted 
in ~ 1998 (while Higgs were found in 2012) with the at that time to the 
129.4 GeV $\pm$ 2 GeV corresponding 
value 135 GeV $\pm$ 10 GeV, when the top mass were a bit higher experimentally 
and accurracy of calculation less, partly behind the head of Mogens Lykketoft
(presently leader of the Danish Parlament): 
\begin{center}
\includegraphics{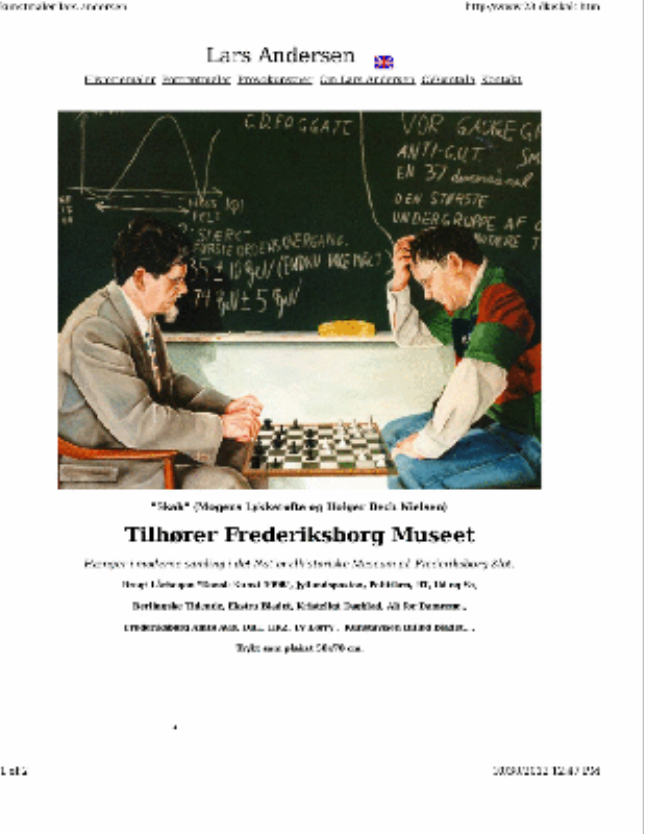}
\end{center}
)

In the following section \ref{support}
we shall put forward - as already 
mentioned above - some propaganda for 
the validity of the ``Multiple Point 
Principle'', which we just saw has got
support from the experimental value 
of the Higgs mass. One of the arguments
is somewhat new, while the other ones are 
reviews. 
In section \ref{counter} we
shall list the arguments for, that we can 
escape using new physics to solve some 
of the otherwise considered as obvious 
problems suggesting  the need for new 
physics.
As subsections we treat  first of 
all the problem of dark matter 
\ref{darkmatter}, but also the usual
problem of the quadratic divergences in 
the in the Higgs mass square is discussed
in the subsection \ref{hp}.
In sucsection \ref{inflaton} we discuss
the suggestion of the Higgs field being 
the inflaton field.

In section \ref{seesaw} we relatively 
shortly call the attention to that 
although our picture is almost without 
new physics untill the Planck scale, we 
admit that we cannot explain the neutrino
oscillations without some new physics, 
which of course typically is a set 
of see-saw neutrinoes. In addition 
to solving the neutrino-oscillations 
\ref{no} we also expect the problem 
of getting an excess baryon number to 
at the end come from the see-saw 
neutrinoes \ref{bn}. 

In section \ref{consolation} we seek 
to consolate the high energy physicist 
reader, who would find it sad, that there 
should be so little new physics.
Finally we conclude in section 
\ref{conclusion} 

\section{Propaganda for Multiple 
Point 
Principle}
\label{support}
In this section we shall put up
some arguments that the assumption of 
``multiple point principle'' of having 
several vacua with essentially the same 
energy density is not so arbitrary as 
it may seem at first. 

Historically we started\cite{MPP} by 
fitting fine 
structure constants in a somewhat special
Random Dynamics\cite{RD} inspired model 
AGUT, in 
which the the gauge groups of the Standard
Model appear as {\em diagonal subgroups
of cross products of several } - one 
for each family - {\em ismorphic groups}.
The crucial assumption to obtain numbers
for the gauge couplings were, that we for 
the generation associated groups took the 
``critical'' couplings meaning lattice 
artifact phase transition couplings. 
It is of course this suggestion of the 
true couplings being just at the phase 
transition point that is essentially 
equivalent to the ``multiple point 
principle''
assumption as we now call it. This fitting
using the AGUT model\cite{AGUT} with the 
lattice 
artifact coupling constants were so 
successfull, that it PREdicted the number
of families to be three! It is not 
surprising, that in our AGUT model with 
its number of cross product factors put 
equal to the number of families and 
a long renorm group running from the 
Planck scale down to the experimentally
accessible scales with beta-functions 
depeding on the number of families
the predictions of the fine structure 
constants at the experimentally accessible
scales become familily-number dependent.
Indeed the two effects of family number
dependence - the number of cross product
factors and the renorm group running -
happened to add up and the number of 
families got fixed by requiring the 
fit of the observed fine structure 
constants. It turned out the fit 
needed {\em three} families in a time, 
when 
there could still have been more families,
because the LEP-experiment measuring 
the number of neutrino species 
\cite{LEP3families} had not 
yet been performed, when we worked on this 
AGUT critical coupling model \cite{AGUT}.

\subsection{Extension of the 
Cosmolgical Constant Problem } 
May be one of best - by words - argument 
in favor of the ``Multiple Point 
Principle'' could be one suggested to 
me by L. Susskind \cite{privatSusskind}. 
This argument goes as follows:

We must accept that the cosmological 
constant is extremely small
\cite{WeinbergCC}  - even of 
order of 3/4 of the present energy density
as astronomically estimated is actually
from all field theoretical points of 
view extremely small - and thus we 
must say ``There exist at least one 
vacuum, namely the present one, which 
has an extremely small -essentially $0$ -
energy density. ''  Now from an estetical
point of view we shall make the most 
simple and beautifull assumtions; that 
is what should be good science to make the
most simple and beautifull model or theory.
Now the point is, that we can so to speak
formulate the assumption of the vacuum 
energy density {\em either in plural or 
in singular}. That is to say we have 
a choice between assuming either that 
there are {\em several} vacua having 
extremely 
small energy density(=cosmological 
constant) - this is the plural version -
or we can assume that there is only 
{\em one} vacuum, namely the present 
one, that has extremely small energy
density(=cosmological constant) -
this is the singular assumption-.
Both these assumptions, singular and 
plural
ones, are about equally estetically 
beautifull. So ignoring our own 
claims that there {\em is} evidence 
for ``Multiple Point Principle''
we only know of one vacuum with 
extremely small cosmological constant,
namely the present one, and it would be 
fifty fifty whether we should believe
the plural or the singular version of 
postulating the cosmolgical 
constant{\em(s)} to be small. 
If we choose the ``plural version''
we have basically assumed ``Multiple
Point Principle''! In this sense 
it would {\em not cost anything 
in terms of complicating the system of 
assumptions} to assume ``Multiple 
Point Principle'' in the form, that there 
are several vacua with extremely small
energy density. 

The figure \ref{quantorsfig} illustrates 
the two versions
of the assumption of the cosmologically 
constant being small
as being approximately 
two versions with different quantors 
in the mathematical sense. In fact 
the ``singular version'' can be considered
described by an existence quantor,
there exist a vacum with extremely 
small energy density, while the 
``plural version'' is more like 
an all-quantor statement saying that 
all the vacua have extremely small 
energy densities. Really of course 
the concept of ``all vacua'' is not 
so clear because  even what a ``vacuum''
is - if it is not just the  grouond 
state - is not
so clear. We might call any minimum 
in the effective potential for the scalar
fields -effective or fundametal - in the
theory  a vacuum, but it is better not
to have to talk about ``all vacua'',
but rather to talk as we say that 
there shall be ``several'' vacua with
same energy density.

\begin{center}    
\includegraphics{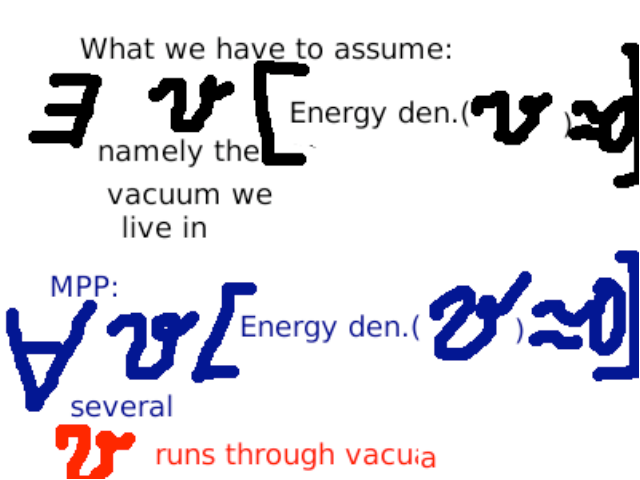}
\label{quantorsfig}
\end{center}

\subsection{Asking for a Model for 
Couplings and Parameters }

One, perhaps almost the simplest model
leading to the ``Multiple Point 
Principle'', takes as its starting point
a wish for constructing a model from which
at least in principle the parameters such 
as coupling constants get determined. That
is to say that we ask ourselves: Can we 
make a model from which the parameters 
can be determined? How to make such a 
model in a not too difficult way to 
construct?

Then the next thought in the development
is to consider the problem that to avoid 
the Hamiltonian 
from loosing its bottom seems very 
difficult. Indeed with renormalization 
and loop corrections it could be very 
hard for the by the hoped for model
predicted parameters to ensure the 
positivity or bottomlessness of the 
Hamiltonian.   It is indeed very hard
even to compute the energy density of
the vacuum of the theory to see,
if the vacuum has e.g. no bottom 
or a negative energy density value. Thus 
it seems 
even harder to produce a mathematical 
formula ensuring the positivity say 
of such a vacuum energy density.
A thought that almost makes it a 
proof that it is impossible to  gaurantee 
the non-negativity of the Hamiltonian 
 eigenvalues is the following:
The  machinery or formula 
to predict the coupling constants and 
other parameters such as masses, upon  
which we think, would a priori be expected
to produce values for the {\em bare} 
parameters. But if so it will depend 
on the cut off whether the lowest energy
density is negative or not. That then 
would imply that the formulas for the 
parameters in the parameter-predicting 
model hoped for would have to be 
{\em cut off dependent}. It  seems 
quite hard to even imagine how that could 
be arranged.

Well, I must  admit that susy
\cite{susy} 
(without supergravity  though) provides 
in a very elegant way a non-negative 
Hamiltonian still leaving the freedom
of a lot of possibilities for the 
coupling constants, that could then be 
fixed by some clever law for couplings. 
But let us 
in the present article ignore Susy as the 
solution, since it is in any case at
least {\em broken}. 

This series of thought is then here 
suggested to encourage to propose a 
model for specifying the parameters 
to contain in its formulation explicitely
the requirement of the positivity of the 
energy density in the theory with the 
parameters being predicted. That is 
to say we simply make it part of the model,
that we state that it must only lead 
to positive or zero Hamiltonian energy 
density.
Now it is normally to be expected that 
such a requirement of say positive 
vacuum or ground state energy density
will lead to {\em inequalities} (not 
equalities)
for the parmeters. Therefore in the space
of possible parameter-values the region
leading to the positivity of the 
Hamiltonian is expected to be of the 
form of a region with some {\em boarders} 
corresponding to the {\em inequalities}  
to be 
satisfied. But if we from this positivity
(really  non-negativity)
requirement only get such a region, we
need some further assumption, if we want 
to have a model giving {\em  fully 
determined 
values} of the parameters.    

But having the requirement of positive 
enegry density ground state (vacuum)
it is very difficult to even 
invent what extra assumption to make 
{\em without spoiling the positivity 
of vacuum energy density}. We so to speak
have to make an assumption that specifies 
the parameters(coupling constants and 
masses) in 
such a way that the specified parameters
are ensured to lie in the region in which 
the vacuum energy is positive or zero.
If we just made some formula for the 
to be specified parameters by phantasy 
or some estetic principe of simplicity 
or the like, we would almost certainly
come to a specification of the parameters
lying outside the by positivity required
region. There is, however, one way to 
propose the specification still 
ensuring the combination of parameters
to lie in the posivity region, namely 
to specify it by a {\em minimization 
requirement}.
That is to say we propose to write down 
by some estetic choice say a quantity 
- a function- of the parameters, call it 
say $S(parameters)$, and then assume:

The specified parameters - by the theory
proposed meant to be the realized values 
of the parameters in nature - shall be 
those giving the {\em smallest possible 
value of} $S(parameters)$ {\em under the 
requirement that the vacuum has positive 
or zero energy density}. 

Precisely by
putting in that the minimization only
is performed over the region with 
the positivity ensured makes it a 
trivial consequence, that this positivity
is ensured, an achievement otherwise 
difficult to ensure. 

Now the main point in the present 
subsection is that {\em rather independent
of the detailed form of the ``estetically
to be chosen function $S(parameters)$''
we obtain the ``Multiple Point 
Principle''} almost independent of 
the details, rather only depending on the 
smoothness of the function 
$S(parameters)$. 

The argument for that Multiple Point 
Principle is strongly suggested now goes
by imagining, how the region of the vacuum 
energy density being positive or zero 
will typically look as a multi-edged 
figure with smooth(but curved) faces 
and edges
in the space of the parameters. On the 
figure we see the region in which 
the ground state (or vacuum) has positive
or zero energy density as a multiedged
figure \ref{figmin}, which if there 
were only 
two parameters - but in realistic 
models as the Standard Model say 
e.g. there are rather of the order of 20 -
would be a polygon with though not
straight sides, but rather surrounded 
by smooth curves in stead of straight 
lines. Now it is rather clear, that 
if the to be minimized function 
$S(parameters)$ is smooth compared to 
the size of the smooth-sided polygon,       it will be most likely that the minimum
will be found on the edge of the 
smooth-sided-polygon  (not truly 
a polygon, because the sides are a
bit curved). It is even rather easily
understood, that it is even very likely 
that the $S(parameters)$-minmizing 
point lies in a corner of the 
smooth-sided-polygon, where a couple 
or even more smooth sides meet each 
other. Now each side of the smooth sided
polygon represents that one candidate 
for a ground state - that a bit depending
on the precise values of the parameters
may or may not become the ground state -
reaches zero. But then it means that 
the very likely situation that the minimum
lies in a corner, where several smooth 
sides cross has the consequence that 
several {\em vacuum candidate are 
just on the boarder of having positive 
energy density by the energy density being
zero}. But that is precisely the situation,
which we called ``Multiple Point 
Principle''. Thus we see that with the 
almost needed type of theory of parameters,
if we shall be able to have a model of 
parameters, we are forced to a type 
of scheme leading to ``Multiple Point
Principle''. In other words: Hoping for
a model delivering the parameters or 
coupling constants in a definite way 
without risking to obtain a negative 
energy density state, leads very 
suggestivly to the ``Multiple Point 
Principle''.

\begin{center}
\includegraphics{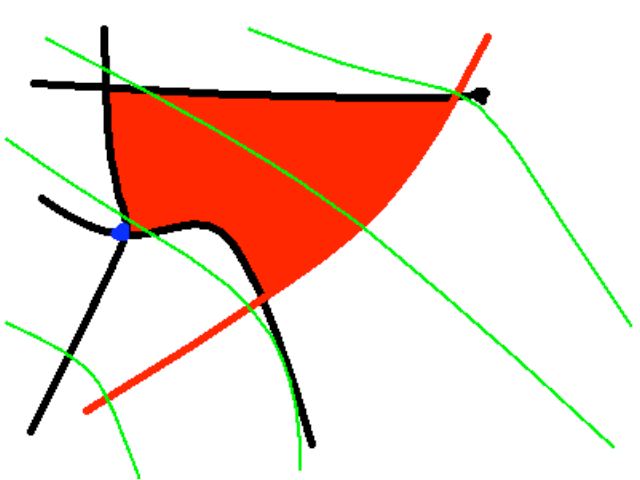}
\label{figmin}
\end{center}

On the figure \ref{figmin} the thin(green) curves 
are the cote-curves for the quantity
$S(parameters)$ which according to 
the imagined model for calculating 
the coupling constants or parameters
should be minimized. The combination 
of parameters, which are realized 
according to the model, should be the 
one inside the dark(red) region 
lying on the curve in the bundle of 
these thin curves corresponding to the 
smallest value of $S(parameters)$ 
among those crossing
at all the allowed dark(red) reginon in
which the vacuum has non-negative energy 
density.    

\subsection{An Example of a to be 
Minimized Quantity an Imaginary 
Part of the Action}  

I have in fact in earlier Bled Proceedings
together with M. Ninomiya\cite{ImSBled} 
put forward
an example of the just discussed type
of model, namely the idea that the 
quantity $S(parameters)$ is indeed the
imaginary part of the action $S_I(path)$.
This is of course then only possible in 
a theory in which the action is not 
real, as it is seemingly observed to be,
but in fact assumed fundamentally
to be complex. To some extend developped 
from earlier works speculating on 
the future influences the past
\cite{old} Ninomiya and I 
\cite{ownmMPP}\cite{ImSBled},
proposed indeed the idea that 
fundamentally the action to be used in 
the Feynman-Dirac-Wentzel path integral
\cite{FDW} for the devlopment of the world 
should be {\em complex } rather than 
real as usually assumed. In this model 
 one obtains by including the future 
(as existing) in the path-integral
that the development of the history 
of the Universe gets determined by 
minimizing the imaginary part of the 
action $S_I(path)$. It is thus first 
of all the history of the Universe
that in this complex action model 
gets determined from the minmization 
of $S_I(path)$, but it may very easily
some way or another get extended to 
also effecively minimizing over different
value combinations of parameters (
$\approx$ coupling constants). Thus 
this complex action theory of ours 
has  indeed very easlily as a consequence
that the complex action $S_I(path)$ 
comes to function as the to be minimized 
quantity $S(parameters)$. Thus in fact 
following the argumentation above we could
check that indeed\cite{ImSBled}  the complex action
theory would lead to the multiple point
principle. 

The most detailed speculation for what 
the imaginary part of the action could 
be is the space-time integral of the 
{\em square of the Higgs field} 
$\int |\phi_h(x)|^2d^4x$, which should 
thus be approximately the quantity to 
be minimized - i.e. S(parameters)- which
though now also depend on what happens in
the world -. That such a minimization 
of the Higgs field squared and integrated 
over the full space-time manifold has 
given a couple of successfull relations 
between indeed couplings, which I 
presented in the article
\cite{phi2minalone}.  

\subsection{Wellness for Human 
Beings
Another Possibility for the 
to be Minimized Quantity, 
Antropic Principle}
It should also not be difficult to come
through with the suggestion that the 
quantity
to be minimized above $S(parameters)$
could be some measure for the chance
of human or human like beings being 
able to develop and survive. Indeed we
must of course imagine that for each 
thinkable combination of the parameters
there is some in principle calculable 
chance that human beings or the like 
develops and become to some status like 
ours, so that they say can think about
coupling constant parameters and measure 
them. If we think of a model in which 
we a priori get a lot of universes created
with essentially all the different values
of the ``parameters'' (coupling constants)
being tested off, we would argue that 
our chanse to live in a given one of 
these being tested off models would 
be proportional to the chanse that 
that model could lead to human-like beings.To be a bit concrete you might think 
of the number of planets which can get 
life depends on some parameter(s) to 
go into the construction of the to 
simulate an antropic principle model 
pratical $S(parameters)$. 

Our point here is that {\em the antropic
principle}, saying that we assume that 
we must exist to observe the combination 
of parameters to being tested, in fact 
{\em is} an example of a model leading 
to fixing parameters of the form we 
suggested, with energy density kept 
essentially non-zero and minimizing 
something, which is then called 
$S(parameters)$. This in turn then 
means according to the above discussion
that {\em antropic principle leads to 
the ``multiple point principle'' very 
likely.}       

    \subsection{Fixing Extensive 
Quantities gives MPP}

The original type of model with which 
we - at first Don Bennett and 
me\cite{MPP} -
hoped to justify the ``Multiple Point
Principle'' was an anlogy to e.g. the 
micro-canonical ensemble in which for 
instance the energy of a system is fixed.
Such a system with an extensive quantity
like energy being fixed will often if 
there is a (first order) phase transition
show up as a system with two phases 
coexisting. Typically one may in 
thermo-dynamics 
ask for the properties of 
some matter under conditions when 
{\em intensive } quantities such as 
temperature $T$ pressure $p$ and chemical
potentials are fixed, and that leads to 
a single phase. However, one can also 
realize experimental situations in 
which one rather have fixed 
{\em extensive} quantitiessuch as 
the volume, the energy, the amount of 
mols of the various types of
molecules, or atoms. The typical example 
of the type with fixed extensive 
quantities - often cited by C. Froggatt
as an introduction of ``multiple point
principle'' - is a botle with stif 
walls, so that the volume of the content
is fixed with a fixed amount of 
water-molecules and further with a given 
energy. One may  keep it termally isolated 
so that no energy can escape or enter
as heat. Then the water in this bottle
is kept with the three extensive 
quantities volume, amounts of mols of 
water and energy fixed. If they happen
not to be fixed to a combination 
of values matching a single phase
of the water there will neccessarily 
coexist several phases. It is actually
easy  without finetuning any of the 
specified extensive quantities to 
arrange that even three different 
phases are required, so that the bottle
will have to contain {\em fluid water,
vapor, and ice} together. If so arranged
the temperature and pressure must be 
at the {\em triple point}. It should be 
stressed that there is a wide range of 
values of the three specified extensive 
quantities leading precisely to this
triple point $(p,T)$ combination. 
In this way it is exceptionally easy 
to arrange the triple point combination 
of the intensive quantities pressure and 
temperature. That is of course really
why it has been so popular to use this 
triple point or other phase transitions 
to define the temperature scale(s).
In the Celsius scale $0^0$ and $100^0$
are respectively the ice to fluid water 
and fluid water to vapor phase transition 
temperatures assumming the intensive 
quantity pressure to be one atmosphere.
But now  a days one rather would use the 
triple point and correct for pressure
going from there to the one atmosphere.

 The reader can  see that there must
be an especially high chanse for in nature
to find such phase transition values 
of the intensive variables compared to 
those intensive variable values not 
connected to any phase transition, the 
latter will namely only be realized 
when quite by accident possibly the  
extensive variable values occur, while 
the phase transition values of  intensive
variables occur for {\em whole ranges 
of extensive
variables}. 

The analogy which we need to this game 
of extensive versus intensive variables
in order to derive the Multiple Point 
Principle is this:
\begin{itemize}
\item To the extensive quantites in 
termodynamics correspond some integrals 
over all space and all time (including
both future and past) of potential 
quantities for being lagrange densities.
That is to say: for each term we could 
think of having in the Lagrangian density
such a $\lambda |\phi_H|^4/4$ we can 
construct an anlogy to the extensive 
quantity as the four dimensional integral
 over all space and all time 
of that quantity, i.e. e.g. 
$\int \lambda |\phi_H|^4/4$. 
That is to say we shall to our derivation 
of MPP use to specify such integrals.
(In the beginning Don Bennett talked 
about ``commodities'' for what has here
been denoted ``extensive variables''
or the integrals over space time).    
\item Corresponding to the intensive
quantities we shall let correspond the 
coupling constants or coefficients rather
in the lagrangian density of the true 
Langrangian for nature.     
\end{itemize}
   
Then we must have in mind a model of the 
type that the Feynman Dirac Wentzel 
path way integral\cite{FDW} for describing the 
be replaced by or better 
supplemented by  some fixation 
of various  integrals over quantities
${\cal L}_j(x)$ that w.r.t. to their 
symmetries could have been Lagrangian 
densitites. In other words 
quantities like $\int {\cal L}_j d^4x$,
which are analogous to extensive quatities
or commodities, must be restricted by 
equations like 
\begin{equation}
{\cal L}_j d^4x = given_j,
\end{equation} 
  where the quantites $given_j$ are so 
to speak God given.

At the end it is then expected that these 
restrictions lead to effectively 
modify the the Lagrangian density to 
an effective one with chnaged coupling 
constants. In this way the coupling 
constant will get effective values, which
 then are the ones for which the 
``Mulitple 
Point Principles'' are derived to work.

It should be remarked that such fixation
of integrals involving both future and 
past times, as we here have in mind, 
means that one has given up the idea that 
the future is absolutely forbidden from 
influencing the past. At least the 
coupling constants - which are the same 
at all times- get influenced by eras in 
time which at some moment were in the 
future w.r.t. that momnet.

But that something like such an influence
of the future on  the parameters or 
couplings is rather unavoidable anyway
were pointed out in article by Don Bennett
and myself: The cosmological constant 
were small with an accuracy that had 
it much   smaller than the energy density 
at the early times shortly after 
big bang. It hardly imaginable that 
the physics in such an early era could 
ever make a so small cosmological constant.
Only knowledge from future in which the 
other energy density sources are of 
the same order or smaller could more 
comfortably be assumed to have an 
influence on the cosmolgical constant.  

\subsection{Mild Non-locallity}
Once you give up strict locallity
in time - which seems hard to avoid for
parameters if the cosmolgical constant
problem shall be solved - and instead 
used say a Feynman-Dirac-Wentzel path 
way integral\cite{FDW} formulation with an only 
mildly local action the ``multiple point
principle'' very easily appear. By this 
mild non-locallity is meant that the 
action is taken to be a function of 
several
integrals over all space of a similar 
nature as what is usually the action,
\begin{equation}
S_{nl}[path] = F(\int {\cal L}_1(x)d^4x, 
\int {\cal L}_2(x)d^4x,...,
\int {\cal L}_j(x)d^4x,...,
\int {\cal L}_M(x)d^4x).
\end{equation} 
 Such an only mildly non-local theory
effectively will appear as local in 
practice, but the coupling constants
will depend on what goes on at almost
all other space and time events, and thus
the only practical non-local effect is 
via the coupling constant. 
This kind model indeed leads to mutiple
point principle. This kind of model 
is described in the cand. scient. thesis
of Nicolai Stillits \cite{Stillits} under 
my advisership.
 
\section{Can we have Standard 
Model All
the Way Up ?}
\label{SMatw}
\subsection{The Suggestion for Only 
Standard Model almost All the 
Way Up}
Let us again mention that taking seriously
our prediction of the Higgs mass on the 
assumtion of multiple point principle
with a competing vacuum with an 
extremely high Higgs field of the 
order of $10^{18}$ GeV not so far from
the Planck energy scale we need to 
assume that the used Standard Model 
to be indeed valid also for such 
a very high Higgs-field $\phi_h(x)$. 

It may of course be that some ``new
physics'' compared to the Standard Model
may only modify the Higgs mass prdiction
of ours so little that it does not matter,
but basically {\em claiming that our 
Higgs mass prediction were not accidental,
then one must have  the Standard 
Model working well at the energy scales 
involved, i.e. in the case energies 
of the order $10^{18}$ GeV.}

This may then have drastic consequences 
w.r.t. what we shall expect there to 
be of ``new physics''.  

\subsection{But is it Not Impossible 
to have Standard Model All the 
Way Up?}

In fact  most high energy physicists would 
believe though to 
find/know several arguments, that the 
Standard Model 
could not possible work almost all the 
way to the Planck 
scale. However, we shall in the present 
article call attention to some of the 
earlier works by the 
present author and collaborators, which 
open
the possibility that indeed there are no 
large deviations 
from the Standard Model almost all the 
way up to the 
Planck scale!

Most importantly we have a somewhat 
speculative and 
also somewhat complicated - but we would 
say not totally 
excluded - picture for dark matter being 
pea size balls
of essentially small white dwarfs having 
inside 
a buble of a vacuum of 
a third type, namely a phase with a boson 
condensate 
of a speculated bound state of 6t and 6 
$\hat{t}$. We shall return to this 
Froggatt's
and mine model for dark matter alone based
on the Standard Model in section 
\ref{darkmatter} below.     
     
Another deviation from the Standard Model 
is the non-zero
masses for the neutrinoes observed via 
neutrino oscillations;
they could for instance be explained as 
due to see-saw 
neutrinoes with some mass, which is high 
compared to the 
presently known particles, but light 
compared to the Planck 
scale. These see-saw neutrinoes may 
couple so weakly 
that they will not disturb significantly 
the running of 
couplings etc. in the Standard Model, so 
that indeed say the fact  
that the Higgs mass should still be on 
the meta-stability or stability
border will not be changed significantly. 
So there would 
be no  argument against such  a weakly 
coupling set of see-saw neutrinoes,
and so there could with  
any realistic measurement accuracy of the 
Higgs mass still be 
 remarkable agreement 
with meta stability or stability limit 
being just 
realized.
This means actually that in the picture 
suggested in the present article the 
see saw neutrinoes make up the first 
and essentially - i.e. untill the 
Planck scale where anyway also 
we expect a lot of new physics - the 
{\em only} new physics. 

Another argument for the need for new 
physics is the hierarchyproblem or 
better the 
associated scale problem of, why the weak 
scale is so low - only say 100 GeV -
compared to the presumably fundamental 
scale, the Planck scale, of energy
of the order $2 * 10^{19}$ GeV ? 
Concerning this point the philosophy 
of the present article is that we 
must {\em postulate some fine tuning 
principle} that specifies the values 
of coupling constants and mass parameters.
One possibility is our ``multiple point 
principle'' which postulates that the 
various couplings etc. get adjusted in 
such a way that a series of different 
vaccuum states all shall be either 
degenerate in energy density, or that 
some vacuum is just on the borderline 
of being metastable. In fact the starting
point for the present article was that 
the Higgs mass were just 
observed to be just on the borderline
for stability of the present vacuum,
so that indeed we started by the 
observation that the ``multiple point 
principle'' (perhaps a bit in the 
direction of the metastability version)
were 
confirmed. We have indeed already 
published a work showing that the multiple
point principle can explain the smallness 
of the weak scale compared to some 
more fundamental scale identified 
with the Planck scale \cite{hierarchybound}.
We shall return to this in section
\ref{hp}.

Yet a problem with having pure standard 
model apart from a bit of see-saw 
neutrinoes all the way to the Planck scale
is the problem of getting sufficiently 
large
excess of baryons over anti-baryons.
In fact in pure Standard Model anomalies 
will at high temperature make the baryon 
number become washed out, only $B-L$ (i.e.
the baryon number $B$ minus the lepton 
number $L$)
would be conserved. So 
unless one has in advance an appropriate
$B-L$ 
there 
would not be a baryon number agreeing 
with the number fitted to Big Bang nuclear
synthesis and to astronomical data. 
However, it is at least possible that 
some see-saw neutrinoes could deliver 
a $B-L$ so that the baryon number fitting
data could be achieved without further 
new physics  below Planck scale than 
the one we have here suggested
\cite{MetaMPP}.

Finally the inflation time in cosmology 
seems to require some new physics, but 
it has been attempted to use the Higgs 
field as the inflaton field, e.g. by 
Kehagias and Germani
\cite{KorfuHi1}. 
Since there 
is for practically any sensible quantum 
field theory with reasonable size of the 
field $|\phi{INFLATON}|$(< Planck scale) 
impossible to 
organize the needed slow roll
\cite{slowroll}, it is 
also hard for the Higgs field to achieve
that. A priori thus essentially all
fields are out of use, not only the Higgs 
field. If - as we shall below
in section \ref{inflaton}   - call for 
some 
``miraculous'' fine tuning help to solve 
the 
slow roll and may consider the slow roll 
problem pushed out, then may be the Higgs
field is not much worse than any other 
field, so that we are no worse off with 
the 
Higgs than with any other field.
\section{Arguing against the 
arguments for
new physics}
\label{counter}
We shall go a bit more in details with 
the arguments for, that it is indeed 
-surprisingly as it may seem to many
colleagues - possible that the 
Standard Model would work well much 
higher up in energy than what is usually 
assumed
and rather apart from a few species 
of right hand neutrinoes, which are 
almost to be considered part of a Standard
Model interpreted slightly liberally, 
work all the way to one or two orders 
of magnitude below the Planck scale.
This may be close to what is called the
Rosner-Bjorken nightmare\cite{Bj}, 
since it at
first seems to be a nightmare for 
physicists hoping to discover new 
fundamental particle species for each 
generation of accellerators. As I shall 
attempt to consolate a little bit below:
the multiple point principle, if it then 
to make up for it were true, could help to 
study by indirect calculations physics at 
very high energy scales via some precision
determinations of couplings at lower 
scales.
   
\subsection{Dark Matter as Balls with 
a Different Vacuum}
\label{darkmatter}
One of the seemingly most obvious
arguments for, that there must be some 
new physics, that can/must even deliver 
copious amounts of matter for the universe,
is the astronomical knowledge of the 
existence of {\em dark matter}. Very 
likely the dark matter could be some new 
physics
particle, which because of some 
conservation such as R-parity in SUSY
models would be so accurately conserved
that these particles could survive the 
13.6 milliard years up to today
\cite{darksusy}. However,
if we now here want to claim that we want 
to obtain dark matter {\em alone} from 
the Standard Model, we have to suggest 
a mechanism for obtaining some objects, 
that can be fundamental particles or 
more or less complicated bound state 
constructions\cite{Tunguska,nbs}, which 
shall be 
stable 
by some mechanism and can be produced 
in sufficient amounts in some rather 
early era. We already know of course 
the practically conserved quantities
in the Standard Model,are  the gauge 
charges,
baryonnumber $B$ and three types of lepton 
numbers$(L_e, L_{\mu}, L_{\tau})$ (with 
sum $L=L_e + L_{\mu}+ L_{\tau}$ ) . The 
latter are then broken by 
the neutrino oscillation effects though.

The model by C.D. Froggatt and myself 
\cite{dark,nbs} shall actually use 
the baryon 
and lepton numbers, in a very similar way 
to how the ordinary matter is stabilized. 
In fact our model 
for dark matter alone with the Standard 
Model should rather be described by 
saying, that we instead of inventing a 
new type of matter to be the dark matter,
invent a mechnism to {\em pack together 
ordinary matter} into packets pressed so 
strongly
together, that the matter packed into these
packets becomes practically so isloated 
from the rest of the ordinary matter 
and from interaction with light etc., 
that our packets can function effectively
as were it a completely different sort of 
matter. It is not difficult to understand,
that, if we indeed can find a method to 
concentrate some ordinary matter into our 
suggested pea-sized balls with weight 
of the order of say $10^9 \ kg$, then  the 
baryons and electrons will occur in such
tight collections, that the interaction 
in normal way of the outermost layer 
of atoms will be so small compared to 
the gravitational force (which is not 
screened in the same way) that such 
balls practically
must count as dark matter.
 
In spite of dark matter being of a density 
bigger than the usual matter density
by a factor of the order 6 in the universe
and say 2 in 
our galaxy the  distances in 
the galaxy between our balls or perls 
is of the order of an 
astronomical unit 

But now, how shall we make such peas of 
concentrated ordinary matter get pressed 
together, so that they can function  
effectively as dark matter? The idea is to 
postulate the existence - which we must 
then in principle confirm by calculation -
of a new phase of the vacuum, in which 
the nucleons obtain a slightly 
(say 10 Mev) smaller mass, so that 
nucleons can be kept  inside this 
new phase. If this new phase is - as 
we shall suggest in our model - mainly 
involved with top-quarks and Higgses 
and as a typical energy scale given 
by the {\em weak energy scale}, the 
tension 
in the walls seperating the different 
vacua would be given
by  dimensional arguments by this 
{\em weak scale}.
If it were not for the special assumption
of ``multiple point principle'' ensuring 
the energy density in the two phases to 
be essentially the same, we would also 
expect the difference in energy density 
between the two phases distiguished 
to be given by the weak energy scale.
If indeed there were such an energy 
density 
difference of the order $~E_{weak}^4$
it would be hopeless for matter made from 
nucleons only changing their mass 
by $~ 10$ MeV by going from one phase
to the other one to stabilize any balls
of an alternaive phase. 
Even the seperation wall
tension threadens to quenche a bit of 
matter made from nucleons if they can 
pass the wall by being pushed just by 10 
MeV energy. When it is about the wall 
pressing the  matter filling a ball 
of a sepearate vacuum, the pressure 
can, however, be made smaller by making 
the 
size of the ball bigger.    
 
So we must 
imagine the balls sufficiently large
- and with sufficiently accurately 
same energy density as the outside vacuum 
- that they do not press out nucleons
which can escape by just an extra energy 
of the order of 10 MeV.
Taking the balls of the bound state 
containing (new) vacuum to be of the size 
of order of 3mm in  diameter the pressure 
we estimate comes down, so that 10 MeV 
mass differnce for nucleons can just 
barely keep the matter inside the balls. 

For being allowed to just think of the 
pressure comming from the wall surrounding
the ball rather than from the difference 
in energy density i the two vacua it 
is crucial, that this difference is 
assumed
effectively zero as multiple point 
princple should ensure. 

Really extensive - even only for 
millimeters - regions with different 
vacua have no chanse to exist were it 
not for a ``multiple point principle''-
assumption, so MPP is crucial for such 
a dark matter model as ours to have a 
chanse. 
But this is what  our whole article 
is based on: the idea of the multiple 
point principle saying that several 
vacua have 
essentially the same energy densities, so
it becomes justified using this principle 
to
assume that the new vacuum has got its 
energy density finetuned to be the same 
as in the usual vacuum. In this way 
we assume away the major cause of pressure
that would have quenched the ordinary 
matter inside the balls only kept by 
a 10 MeV potential. There will, however, 
expectedly exist a wall between the 
two phases - the new and the usual 
phases - and that obtains from dimensional
arguments reason a tension given also
from the weak scale energy say 
$E_{weak}~100 GeV$.
In order that this tension shall not 
by itself lead to pressure bigger than 
the 10 MeV per nucleon as we imagine the 
matter could stand before getting pushed 
out, we need rather big balls. It is in 
this way we reach to the ``pea-size''
proposed. The surface tension of the 
wall around a ball taken to be of the 
weak order of magnitude $tension \approx 
(100 GeV)^3$ will provide a pressure 
of the order of this number divided by
the radius of the ball. The size of the 
ball which we expect can keep the nucleons
inside it with a potential difference 
of the order of 10 MeV has to be around 
the 
pea-size 1 cm. We imagine the balls pumped
up almost to so high density that the 
ordinary matter is just about to be pushed 
out - the degenerate electrons genuinely
providing the pressure has to go up to
Fermi-sea with a fermi-energy of the order
of the 10 MeV. The situation is pretty
well described as a white dwarf pressed 
together by the wall surrounding the 
pea-size ball of new vacuum. If the ball
is sufficiently big not to collaps, its 
baryon number $B$ and lepton number 
$L \approx L_e$ function 
as 
conserved quantities and the ball will
exist essentially forever. It will be no 
problem to have it existing for the 13.6 
milliard years, in which  the universe 
has existed.

We should at this point remark, that 
our model is superior to the usual type 
of model with susy-partners
\cite{darksusy}, which 
requires 
a special - only by postulate conserved -
quantity (R parity) to explain the 
remarkable stability of dark matter,
which is needed. In our model we 
{\em reuse
the baryon number}, which is already well
understood to be very well conserved in 
the Standard Model. 

But by having dark matter thus basically 
being the same as ordinary matter except
for sitting on an other vacuum it becomes 
very important for that our model 
shall be able to fit 
the big bang nuclear synthesis -  which 
seems to explain well the abundances 
of the lightest isotopes already
at the era of this Big Bang Nuclear 
Sythesis era - that the dark matter is 
well packed 
into the small balls already at that time.
 Otherwise it might
disturb the big bang nuclear sythesis.

\begin{center}
\includegraphics{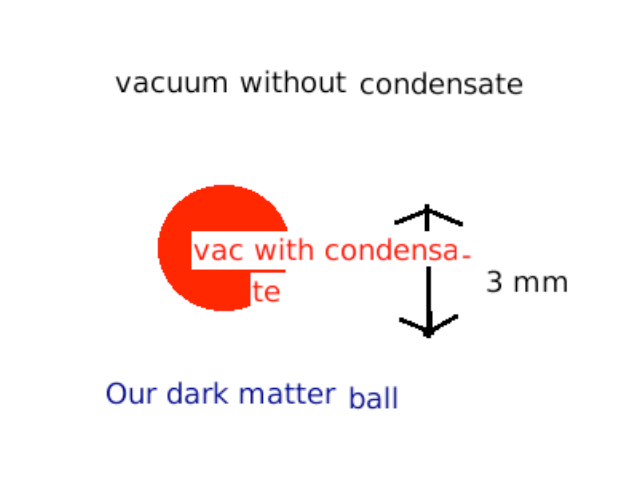}
\end{center}  

\subsubsection{Seeing Dark Matter on 
Earth?}       
In these days it seems that the DAMA 
experiment\cite{DAMA} has 
already {\em seen} dark 
matter hitting the earth in as far as
this experiment has seen a 9 standar 
deviations season variation of potential
dark matter caused events. If these to 
some extend at some time 
seemed 
to disagree with other  observations are 
indeed 
correct and indeed observations of the 
dark matter, our model would be 
disqualified. In fact if the DAMA 
experiments indeed as it seems mean
that dark matter is in the form of 
particles distributed so as to be so 
many particles that they can be observed 
by DAMA then our model of pea sized 
balls is out. Particles to satisfy the 
requirements for being observed by DAMA
will presumably be extremely difficult
to obtain in a pure Standard Model; so 
if DAMA is not explained away somehow,
then the main thesis of the present 
article, that the Standard Model should
work all the way up could hardly be 
uphold. We therefore have to 
hope for that the DAMA experiment can be 
explained away.  Otherwise  we cannot 
 uphold our 
thesis about the dark matter. 

But even if now dark matter were indeed 
our type of balls and the experimental 
observations were somehow a mistake,
we must ask: would the  earth not be 
hit by the dark matter anywhere ?

Indeed we have had success with fitting 
the size of our dark matter balls to such
sizes that they do match the density 
after weight of dark matter from astronomy
will hit the earth about one every hundred
years.  In fact we put forward 
- Colin D. Froggatt and I \cite{Tunguska}-
the idea that the event that happened 
in Tunguska about 100 years ago were 
in fact caused by a dark matter ball
hitting the earth. 

When we say that we fit well with 
our dark matter balls, we mean first
of all that the hitting of one ball 
about every hundred years  or two hundred 
years combined with the knowledge 
of the dark matter density astronomically
estimated in galaxy halo fits well
with the size that from particle theory
estimates is about the minimal size 
that can be stable.

\begin{center}
\includegraphics{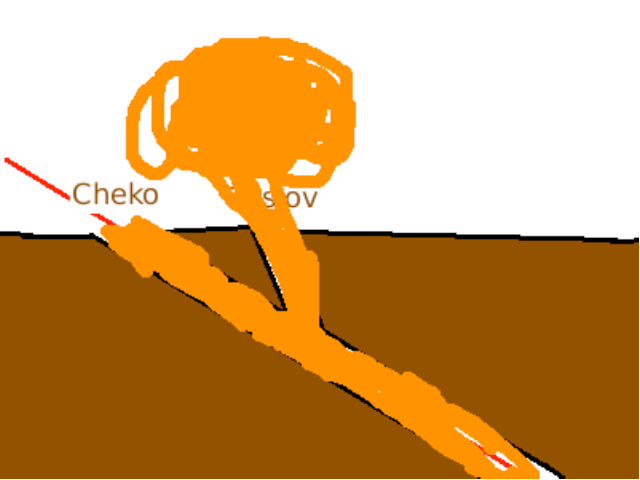}
\end{center}   
\subsection{Hierarchy Problem 
Essentially 
Solved by Explicite Finetuning 
Assumption}
\label{hp}
Now when we take as our great new
physics assumption that we allow ourselves
to finetune - namely via the ``multiple
point peinciple'' assumption, it is of 
course our hope that we do not only 
thereby 
solve the cosmological constant problem
- which is essentially the one we give up 
solving, but just generalize - but also 
the other important fine tuning problem:
Why the weak scale is so enormously low 
compared to the Planck scale?

This we have actually done in earlier 
works together with Larisa Laperashvili 
and  Colin Froggatt\cite{hierarchybound, nbs}. 

The crucial point is that we assume there
to be three essentially zero energy 
density vacua in the Standard Model:
One is the one we live in, the second 
one is the one with the about $10^{18}$
GeV Higgs field, which is relevant for
deducing the Higgs mass, and the third 
one is the one with the condensate of the 
bound states of the 6 top and 6 anti-top,
which is supposedly realized inside 
our dark matter balls. 

We might now put our ``solution'' of the
scale problem just mentioned as follows:

The degeneracy of the one we live in 
and the $10^{18}$ GeV Higgs field one 
implies a value for the {\em running}
top-Yukawa coupling $g_t(t)$ for $t$
at the almost planck scale, because it
is really needed that the running 
selfcoupling $\lambda(t)$ shall not only
be approximately zero at the ``almost
Planck scale'' $10^{18}$ GeV, but also
its derivative w.r.t. to $t$ shall
be zero there in order that there be a 
minimum in the effective potential (which
is approximately given as $V_{eff}(\phi_h) 
=\lambda(\log
(\phi_h))*\phi_h^4/8$ ), i.e. the 
derivative of the effective potential 
$V_{eff}(\phi_h)$ shall be zero. 

This leads assuming the known 
finestructure constants etc. to a 
top Yukawa coupling up there of 
size 0.4. Next the degeneracy of the 
vacuum we live in and the boundstate 
condensate one gives similarly that 
the running top-Yukawa coupling 
must be now at the weak scale 1.02$\pm 14 
\% $ 
as Froggatt and I calculated. This 
latter calculation means that we require 
the top-yukawa coupling to be large enough that the binding of the 6 top + 6 anti top
quarks by mainly Higgs exchange will 
be just so stronmg as to guarantee that 
the binding energy just compensates the 
Einstein energy of the 12 top or anti top
quarks so as to make the bound state 
massless.       

Then the crucial point is that the 
top-Yukawa coupling has to run a 
prescribed amount from .4 to 1.0 in order
to fullfill the multiple point principle!
But now with the typical order of 
magnitude of the couplings in general
this running is so slow that a ``long ''
range on the {\em logarithmic scale} is 
needed.
It turns out that this needed  range 
measured in  the 
logarithm 
is 
very much of the right 
order of magnitude to explain the 
smallness of the weak scale compared to 
the Planck scale! Indeed if one requires 
the experimental top quark Yukawa coupling 
at the weak scale .93 and require it 
to match with the 0.4 at the Planck scale 
one gets very good agreement for the 
weak scale to Planck scale ratio.

In this way we can claim almost to have 
calculated crudely the weak scale to 
Planck scale ratio!

So indeed we can claim that the 
``Multiple Point Principle''  solves 
the scale problem and thereby in a way 
also 
the ``hierarchy problem''.  

To solve - also- the hierarchy problem
one would have to accept that the loop
corrections to Higgs mass square would 
still contain the (in)famous  quadratic
divergences, but that we should 
renormalize to the ``multiple point 
principle'', meaning that we should 
loop correct the bare Higgs mass square 
by a quadratically divergent term
so as to {\em cancel the divergences in 
the 
vacuum energy densities} of the three 
vacua,
since the latter have from ``multiple 
point principle'' to be zero.

That is of course not a true solution 
getting rid of the enourmous quadratic 
divergences, but rather a {\em 
renormalization}
only. But we can renormalize to a 
theoretical principle namely ``multiple
point'' rather than just to experimental
data. Including this ``mutiple point
principle'' in our calculational rules
this renormalization could be considered 
automatic and then the Higgs mass (square)
would not be shuffled around in the usual 
crazy way by additions of huge 
quadratically divergent contributions.
Instead it would remain in the weak 
scale range. For this to be successfull
it is one should though have in mind 
that the non-perturbative calculation 
of which yukawa coupling give the 
binding of the 6 top + 6 anti top just 
to the phase transition should be 
included into the calculation.
One could only go on to do higher and 
higher accuracy and get rid of the 
crazy shuffling around provided one 
in each level of accuracy already has 
a reasonable accuracy for the bound 
state and the phase transition point
(in the top yukawa coupling).    

\subsection{The Inflaton Could be 
a Higgs ?}\label{inflaton}
Although the scale of energy density 
during the inflation period may not be so 
well known, it may be best to take it that
this energy density were several orders 
of magnitude below the Planck energy 
density, and that indeed the energy scale
relevant for inflation is smaller than the
Planck energy scale. An estimate is found 
in\cite{inflationscale} for the typical 
energy for inflation and thereby the 
``reheating'' temperature.
This means, that if we insist on having 
only the Standard Model (though admitting 
hopefully  unimportant sterile 
neutrinoes) ``all the way up to Planck'',
then we should arrange for even the 
inflation to be totaly  understood in 
terms 
of Standard Model physics. Thus assuming 
the inflation to go by means of a scalar 
field resting under the inflation proper 
at some high effective potential value 
and then at reheating time falling down 
to the final ground state, we have only 
the single scalar field in the Standard 
Model, the Higgs field to play with.
Usually it is told that it is impossible 
that the Higgs field can function as 
the inflaton field. That is indeed true
if you note that the Higgs field has the 
peak in its effective potential below 
the   Planck scale field value, and that 
all
potential inflaton fields with their 
plataue or their peak below the Planck 
scale have a ``slow roll'' problem\cite
{slowroll}. I would consider the 
assumption that we should have the 
inflaton field
strength during inflation not be 
many orders of magnitude larger than 
the Planck energy a very reasonable one.
Thus I would take it that all 
``reasonable'' inflaton models should 
work in inflation time at an essentially 
subplanckian point or at least not much
above. Then it is rather easy to argue 
that all reasonable inflaton models 
are unable to produce the say 70 
e-foldings of inflation required for 
the inflation being able to naturally 
bring the energy density sufficiently 
close to the critical density so as 
to ensure that the density of energy 
in the universe will not again move away 
from criticallity (untill today).
 
\subsubsection{A Theorem on Slow Roll}
Let us here state the slow roll problem
in the form of the following theorem:

{\em With a polynomially and 
renormalizablily smooth effetive potential
for the inflaton field a large number 
of e-foldings is not achievable under the 
assumption that the inflaton field 
sejours at a field value not  much bigger 
than the Planck energy scale}

Argument  and explanation: By 
``polynomially 
and renormalizably smooth'' we assume
that in first approximation the effective 
potential has approximately
(renorm group corrections should be
allowed)  the form of 
a polynomial
of the up to fourth order as is allowed 
in the classical approximation for a 
renormalizable theory, but really this 
assumption is not so important.  
Using Planck units makes order of 
magnitudewise simply the Hubble constant 
equal to $\sqrt{V(\phi)}$. Suppose we 
have the expansion going on in the 
inflation time with the approximate 
field value $\phi_{sejourn}$, according 
to the assumption in the theorem being 
less than or about equal to the planck 
energy scale. The second derivative 
at this $\phi_{sejourn}$ value must 
obey order of magnitudewise
\begin{equation}
|\frac{d^2 V(\phi)}{d\phi^2}|_{\phi 
= \phi_{sejourn}}| \ge 
\frac{V(\phi_{sejourn})}{\phi_{sejourn}^2} 
\hbox{(order of magnitudewise)}.
\end{equation}
because the effective potential is 
crudely just a low order polynomial.      
Now we get the from a Taylor expansion
and approximation by an inverse Harmonic 
oscillator obtained time scale for the 
exponential run away from the peak 
value $t_{iho}=1/\sqrt{|
\frac{d^2 V(\phi)}{d\phi^4}|
_{\phi = \phi_{sejourn}}|} $.
This time scale is thus order of 
magnitudewise not larger than 
$1/\sqrt{\frac{V(\phi_{sejourn}}
{\phi_{sejourn}}} = \frac{\phi_{sejourn}}{H}$.
Now we assumed that $\phi_{sejourn}$ 
were not (much) greater than the Planck 
energy $E_{Pl}$ for the moment used as 
unit. Thus we have also derived that 
the time $t_{iho}$ of the inverse harmonic 
oscillator  cannot be greater than 
the inverse Hubble constant at the 
inflation time $1/H$. But this then means 
that the number of e-foldings - of which 
there occurs one per Hubble time 
$1/H$ cannot order of magnitudewise 
be more than of order unity. 

So under 
each e-folding the 
distance between the peak and the 
actual field value grows also by 
a factor of order unity, essentially 
also an e-folding.  

Now in the Planck units, which  we use 
for the 
moment, we can count the $\phi_{sejourn}$
as less than 1 and there is no big 
number involved and we should not 
get very large number out, let alone a 
so large number that even its logarithm 
should be large. (The number we are
to look for is the scale factor under 
inflation, which should at least be
$exp(70)$ say.)   

So we can not obtain a long time inflation
measurede in that time and we can thus 
not get many e-foldings! 

{\em End of argument/proof.}

So one problem with having the Higgs field
being the inflaton field is the slow roll
problem, which is a problem for any 
sensible assuming most importantly that 
the field value during inflation is not
much larger than the Planck scale. 
This of course means that there is no
special reason for excluding especially 
the Higgs field, but of course it means 
that the Higgs does ALSO NOT function as 
the inflaton.

If, however, we accepted to go for 
a yet to be found or in other ways 
outrageous solution to the slow roll 
problem, such a solution might help
also the Higgs field to become after all
a candidate for being the inflaton!
\cite{KorfuHi1}

What I have in mind as a candidate for 
some ``outrageous'' type of solution
to help on the slow roll problem, still
using a quite sensible scalar field theory
and even with the field ranged being used 
lying below the Planck scale field 
strength, would be to used the already 
above mentioned ``complex action model''
by myself and Ninomiya\cite{ownmMPP, 
ImSBled}, in which 
there 
is influence from the future\cite{old}. It is 
logically possible to imagine that 
the imginary  part of the action has 
such an expression a long inflation time 
would be favorable towards minimizing 
this imaginary part $S_I[path]$. Thus 
this influence from the future could 
possibly cause  the 
initial stand of the inflaton field 
to lie so finetuned on a peak of the 
effective potential that is would take
very long, say 70 e-foldings or rather 
70 Hubble times, before the field value
has fallen down toward the  minimum.
In fact K. Nagao and I \cite{NagaoSI} are 
for the time 
being working on a study inside such 
a complex action model on examples 
such as the harmonic oscillator and the 
for the present discussion most important 
example, the inverse harmonic oscillator.
An inverse harmonic oscillator is the 
simplest approximation to a system 
consisting of a single non-relativistic 
particle in one space dimension 
moving in the neighborhood of a maximum 
of the potential. In real quantum 
mechanics the unavoidable uncertainty 
(Heisenberg inequality) puts a limit to 
how long the particle statistically can 
sejourn near the peak, although 
classically a sufficently exact finetuning
could keep it standing arbitrarily long 
time. 
Intuitively we would get surprised 
if we saw a pen, say, standing straight up 
on its tip for a few days, and usually it
does not happen, so either quantum 
mechanics or other effects causes it to
fall even if we have made quite an effort
to make it stand very straight and 
excercised with putting it up so as 
to make it stand long. But in principle
one could imagine some fine tuning 
``complex
action theory'' to deliver a more accurate
finetuning. We should have in mind that 
ignoring or averaging over the spacially
varying fluctuations of the inflaton field
an inflaton starting near a peak in the
effective potential is approximately 
an inverse harmonic oscillator.
But now indeed it looks that our
\cite{NagaoSI} inverse
harmonic oscillator studies point to 
that with future included and in complex 
action model it is indeed possible to 
get in in a likely way a solution 
favoured, 
which is fine tuned! 

If we made use of such a finetune state 
machinery, it would mean that we looked 
for what with some right could be 
called a ``miracle'' called in to solve 
the
slow roll problem. Even if the 
reader should not feel attrackted to such 
a ``complex action model'' even if it 
solved 
the slow roll problem, at least such 
a model would constitute an 
example of  how one in desparation 
could imagine to solve the slow roll 
problem with inflation. That example 
would now work on the idea of having the 
Higgs being the inflaton as well as on 
other proposals. So if we found some 
solution like this, a ``miracle'' solving
slow roll, that in general could solve
that problem, a major reason for the 
Higgs not being, as seen at first, 
a good inflaton would disappear!

The main point of this argument, that 
the need for new scalar fields to 
have the inflation working and thus 
needing at least more than the Standard 
Model at the say reheating energy or 
temperature scale is, that I answer it 
like
this:

It is true, that in the picture of the 
Standard Model all the way up to 
an order of magnitude under the Plack 
scale, except for unimportant right 
handed or see-saw neutrinoes, we have 
{\em only the Higgs field} to play the role
of the inflaton-field. And it is true 
that the Higgs field applied as 
inflaton-field leads to a slow roll 
problem, meaning really that it does 
not work unless somehow helped by 
something quite new. However, having 
instead
a field outside the Standard Model would
not help much, if we kept to a ``sensible''
assumption of not letting the field value 
taken on (under the inflation) be much 
larger than the Planck energy scale, 
because then there would {\em still be 
a slow roll problem}! So I say, if we 
somehow solved the slow roll 
problem for some scalar field in a  
``reasonable'' scenario, then the same
solution might also very likely solve 
it for the Higgs field; so why throw
out the Higgs as a candidate, when 
it is after all likely to be competitive 
with the 
alternative scalar fields, that could 
be added to the Stanard Model. Alternative
fields being of significance at a 
reheating or inflation energy scale 
several orders
of magnitude below the Planck could 
at least in principle disturb our 
Higgs-mass prediction, and thus they 
would not be wellcome in the scenario
of the present article of taking the 
success of this agreement seriously.

\section{Even \underline{We} must allow 
See-saw 
Neutrinoes for: 
Neutrino-Oscillations
And Baryon Number in Universe}
 \label{seesaw}
We must here admit that even WE cannot 
propose that the Standard Model should 
work truly all the way up to the Planck 
scale order of magnitudewise. The reasons 
are the Neutrino-oscillations which 
clearly shows that the lepton numbers
for the seperate families of leptons 
are definitely not exactly conserved.
In the Standard Model we do namely have  
the seperate flavours of lepton numbers,
electronleptonnumber, muonleptonnumber,
and tauleptonnumber, as accidentally 
conserved quantities. Well, we do not
really have  these lepton 
numbers truly conserved, when anomalies 
only 
active at high temperatures 
are counted. Then namely 
it is only the baryon number $B$ minus the 
(total) lepton number $L$ i.e. $B-L$ that is conserved. But
so high temperatures are definitely not 
involved in the neutrino oscillations 
observed. These neutrino oscillations 
are also of a too large order of 
magnitude to be consistent with there
being only new physics at almost the
Planck scale. Thus it is unavoidable 
to have some new physics at lower
scale than the Planck scale in order to 
have the neutrino oscillations as observed.

Similarly we might say that in order to 
get the $B-L$ violation needed for making 
the baryon excess, which is observed, a 
violation of the in the Standard Model 
``accidentally''\footnote{We use the 
terminology ``accidental'' symmetry 
and ``accidentally'' conserved quantity 
for symmetries 
and corresponding Noether charges, when 
they appear in a quantum field theory, in 
which the symmetries directly imposed 
as definiton of the quantum field theory
in question do NOT include that symmetry
or conservation law. It means that the 
``accidental'' symmetry or conservation 
law come out by a slightly more 
detailed looking at the theory, but 
it has {\em  not} been imposed as say 
the gauge 
symmetry of the theory, and it has 
{\em not} just been assumed for Lagrangian
density. You read for 
instance simply off the most general 
renormalizable Lagrangian density 
restricted by the gauge symmetry 
requirements in the Standard Model, that 
baryon number and the 
various seperate lepton numbers 
(electronlepton number muon lepton 
number ..)are conserved. Now the 
anomalies violate
some combinations of these symmetries, 
but e.g. $B-L$ remains conserved even 
when anomalies are included. So one 
might call the by anomaly broken ones
anomaly-broken accidental symmetries,
while $B-L$ is a ''true accidental 
symmetry'' in the Standard Model.} 
conserved
quantity $B-L$  is needed. Here some 
say Majorana see-saw neutrinoes could
do the job. They could namely have 
Majorana-mass terms in the Lagrangian.
(Such Majorana-mass-terms, if they should 
reach the scale suggestive for see-saw 
neutrinoes, would have masses much 
smaller than the Planck scale even if 
very large compared to the weak or strong 
scales say. Thus these Majorana masses
constitue in principle a fine tuning 
problem. Possibly though a relatively
large number of approximately conserved 
quantum numbers being revealed as gauge 
quantum numbers only one order of 
magnitude under the Planck scale, could 
make the suppression of the Majorana
masses sufficient\cite{Yasutaka}).

\subsection{Neutrino-oscillations}
\label{no}
In several articles Yasutaka Takanishi 
and myself and also  
partly in collaboration with 
Colin Froggatt 
have built up a somewhat complicated 
model\cite{Yasutaka} which, however, only 
gets 
complicated and is an extension of 
old models of myself and Brene, Bennett 
etc called AntiGUT\cite{AGUT}, when 
we consider
energy scales close to the Planck scale.
From say about one order of magnitude 
below the Planck scale and lower on 
this type of model has ONLY the seesaw
neutrinoes and the Standard Model 
particles. All the modelling we
speculated were only to fit the detailed 
orders of magnitude of the see-saw 
neutrinoes and the Yukawa couplings 
so as to fit the little hierarchy problem 
and the neutrino oscillations orders of 
magnitudewise. If one accepts, that we 
are not
yet at the stage of being able to fit 
the Yukawa couplings generally, our model
would as the only new physics have the 
see saw - i.e. right handed or better
sterile neutrinoes - in a mass range 
up at $10^9$ to $10^{12}$ GeV or so.

The degree to which such sterile 
neutrinoes will disturb the for us 
so wonderfull prediction of the Higgs mass
is determined from the degree to which
they influence the running of the Higgs 
self coupling $\lambda_{run}(t)$, because 
it is to first approximation this running
selfcoupling, that gives us the effective 
potential for the Higgs field $\phi_h$.

Indeed the effective potential for the 
Higgs field $V_{eff}(\phi_h)$ is because 
the smallness of the Higgs-mass at least
for large Higgsfield values $\phi_h$ 
given approximately alone by the fourth 
order term
\begin{equation}
V_{eff}(\phi_h) \approx \frac{\lambda_{run}(
\mu = \phi_h)}{8}*|\phi_h|^4. 
\end{equation}

The sterile neutrinoes do couple to the 
in low energy physics known ``ordinary''
neutrinoes and a Higgs, and there is thus
basis for that they can provided diagrams
contributing to  Higgs self energy 
as well as to the Higgs self 
interaction and thus to make the running
\cite{beta}
of the self coupling get modified. 

However, if the couplings between the 
Higgs and ordinary to sterile netrino 
transiton are of the ``usual'' small size 
like all Yukawa couplings except for the
top-quark Yukawa coupling their 
contribution will be of a similar order 
of magnitude as those from the
majority of the quarks and leptons, and 
that is of negligible magnitude.

So the only ``danger'' for that this 
sterile neutrino new physics would 
disturb our Higgs mass calculation 
would be if the Higgs coupling for them to
the ordinary neutrinoes would be of order 
unity. If they are suppressed by 
the ``usual'' ``small hierarchy problem''
type of suppression they would only 
disturb very little. So provided 
such a ``usual'' suppression we could
accept such sterile neutrinoes, and 
still have our Higgs mass prediction 
be non-accidental!    

\subsection{Baryon number excess}
\label{bn}
In the see saw neutrino models the 
natural assumption is that the see saw 
neutrinoes at a stage of the cosmological
development cause an over-abundance 
of $B-L$, as an under-abundance of
lepton number $L$ due to lepton non 
conservation
and timereversal assymmetry 
\cite{Sakharov}. 

Our own model \cite{YTCDFYC, Yasutaka} in 
fact shows that it 
is far from unlikely that some model with
sufficiently small Yukawas connecting the 
sterile neutrinoes and ordinary neutrinoes
that they could avoid disturbing 
the self coupling running significantly.
We must however admit that our special 
model then had problems in fitting well
the baryon assymmetry. However, 
that problem came about via a strange 
detail in our model: We actually obtain 
at one moment of the cosmology era a 
good for fitting excess of $B-L$, but then 
we have some rather light sterile 
neutrinoes surviving and having themselves
too little CP-violating couplings so that 
they remove the already produced $B-L$.

This problem is very much a detail 
of our model building and would be 
extremely easily removed, if one just wants
to explain that it is indeed quite likely
that the sterile neutrinoes do not have to
disturb our Higgs mass prediction.
 
\section{Consolation for no new 
physics}\label{consolation} Sometimes 
one hears it as 
a very sad\cite{Bj} happening for the field of high 
energy physics, if it turns out that there
is no ``new physics'' at the LHC scale.
We want here to to some extend argue for 
some consolation in the case suggested 
in the present article, namely that there 
are some relations between the parameters,
i.e. the coupling constants and the 
Higgs mass, of the Standard Model of a
nature requiring the validity of the 
Standard Model up to close to the 
Planck scale. If such relations turned 
out to be true, we could in priciple study 
almost Planck scale physics indirectly 
by measuring and understanding the 
relevant parameters at ``low'', i.e. 
say LHC, energies.  In principle we 
could need more and more accurate 
values for the parameters in order to 
indirectly settle more and more details 
about the for the couplings relevant 
scale, although this scale itself might 
turn out exceedingly hard to truly get 
a direct access to. If we want to get 
more accurate knowledge about the values 
of the parameters when extrapolated to 
close to the Planck scale, then it is 
usefull to obtain the values of these 
parameters at as high energy scales 
as we can get to know them in order to 
have so short distance in scale-ratio 
left to extrapolate to
reach the supposed relevant scale 
relatively close to the Planck scale.   
           
Now it must also be admitted that our 
picture although formally we have only 
the Standard Model does indeed contain 
our proposed bound state of 
$6t + 6\hat{t}$
which one might find experimentally, 
e.g. the Higgs might decay into a pair 
of such bound state particles.
(That seems however not to be 
case, because if indeed the Higgs decayed
into our bound states, then the Higgs 
would decay away and would not have been
observed so far.) In some 
sense finding such particles would really
be an indirect very accurate measurement 
of say the top-quark Yukawa coupling
in a combination with the complicated 
binding mechanism going on. So the 
precission measurement would in this 
special case in fact in pracsis take
the character of finding ``new physics''
much like, if it had been a supersymmetric 
partner. It is just that now in principle
our bound state is fully understandable
at the end in terms of a Standard Model 
story. Very recently we \cite{Larisatalk}
are looking for that the existence of 
a very strongly bound state or of several
such bound states could lead to changes
in the in Standard Model calculate Higgs
production and decya rates, especially 
the Higgs ---> $\gamma + \gamma$ would 
have a sensitive decay rate because it
is already in the Standard Model as 
naively (i.e. without our bound state)
given by loops.

In the next accelerator that might be 
the ILC, the international linear collider,
one might get presumably better accuracy
than in the hadron colliders. So that 
might give better chance for extrapolating
and make a fit to the values of the 
coupling constants, once we may have 
developped some machinery for determining
the coupling constants. 

A priori we would say, that it does not 
matter so much, if the information 
we by the experiments extract out of 
nature to teach us the say Planck scale 
physics comes via an understanding of 
the coupling constants or via seeing 
more particles as one goes along with 
higher accelerators. What should matter 
should rather be how big is the amount 
of information gained\cite{Rugh}. The
crux of the matter is that we have enough
information, enough accuracy of 
parameters or how rich spectrum of 
particles, that we can reach to claim
that a sufficiently complicated theory 
must be right, because it after all 
is no more complicated than that it 
must be right if it can explain a set
of information rich data. We must face
that most likely we should not imagine 
that the final theory will be so simple
that we would have to believe it unless
it can support its truth by a reasonable 
large amount of data fitting. 
If we face that the degree of simplicity 
of the final theory as we shall conceive 
of it is not so great that we can trust
the model immediately, then we must 
have some sufficient amount of also 
rather accurate data in order to 
get such a theory justified. We may simply
stand in a situation, that only if we can 
get the extra accuracy achieved by say the
ILC can we come to great enough accuracy 
to settle if the proposed theory is right.
This may be so even if we should at that
moment be sure that no new physics 
should be immediately found by the next 
machine. 

One shall not be too sad, because we might 
already be so far that we already know 
the theory as relevant crudely at the LHC 
scale. The theory shall also sometimes 
work, otherwise what would be the 
purpose of having a theory?
            
\section{Conclusion }
\label{conclusion} We have pointed out 
that the Higgs mass $126 \ GeV \pm 
~2 \ GeV$ which is slowly 
gotten settled suggests the correctness
of a theory that like the ``multiple 
point principle'' let the coupling 
constants and mass parameters be 
determined by vacuum-properties 
(degeneracy or barely metastability or 
assumptions about the transitions between
the vacua).  In fact the ``multiple 
point principle'' predicts with present
top mass and calculational status
$m_h = 129.4 \pm 2GeV$ (for requiring 
degeneracy). Even more 
important, if this
is taken seriously the theory used to
calculate the involved extra vacuum,
which fits the  Higgs mass found 
experimentally , must really be 
true with the for the prediction 
sufficient accuracy.
In the study here the theory that in this 
way gets suggested to be true 
is {\em the Standard 
Model being valid all the way to almost 
the Planck scale}. We therefore suggest
that indeed the Standard Model shall be 
true 
all the way up, and only new physics of 
an unimportant character can be tolerated
appreciably below the Planck scale.
  
We have then argued that using various 
models of ours etc. in fact the scenario, 
that there be {\em only some see-saw 
neutrinoes} below the Planck scale - 
or better below about one order of 
magnitude below the Planck scale - and 
no other physics in excess to the 
Standard Model, is viable!

If indeed it should turn out that at 
the LHC one  sees no ``new physics''- not
counting the Higgs as new physics, nor
our  proposed bound states, if they 
should exist - and the Higgs mass indeed 
turn out to be the one we predict, then
one would have to take serious that one 
should  
consider our picture. 

At first one of course just think about 
postponing the ``new physics'', containing
SUSY and/or some dark matter candidates of
conventional type, up to a higher energy 
scale so as to get above LHC, in case 
one had seen nothing there, but still 
much lower than the Planck scale. 
Of course such a scenario with postponed 
new physics just a fraction of or a few 
orders of magnitude above LHC is possible 
logically.

However, I would claim that the longer the
new physics is postponed the more the 
finetuning gets called for. That is to say
the more strange it becomes to have 
the tuning in of the Higgs mass scale 
to a value much smaller than the 
SUSY-breaking (if say the new physics were
SUSY). Also the specific value of the 
Higgs mass, and here I think of the 
precise value close to ~$129.4\pm 2 \ GeV$, rather 
than 
just the order of magnitude, would 
stay as unexplained even though this 
value has special significance as the 
minimal possible in the Standard Model
- taken to be valid all the way up - 
supposing (meta-)stability of the existing
vacuum.

So one might argue further in the LHC not 
finding new physics case:

Even to avoid finetuning of the Higgs 
mass quadratic divergences a fine tuning 
is unavoidable. So there would be no way 
to escape fine tuning, and thus the best
would be to look for a {\em finetuning 
law}. That were of course what our 
``multiple point principle''
- or our theory of complex action which
could be considered a model behind 
the `multiple point principle'' - 
precisely is, a {\em law for finetuning}.

Even new physics at a for the moment
unaccessible scale above the LHC energies
could clearly serve as dark matter. We 
know too little about dark matter and 
as our own proposal of pea size balls 
show the energy per particle in the 
dark matter is totally undetermined.

There is also of course no strong reason 
for like here proposed to use the Higgs 
as the inflaton. It is namely  as far as 
it is 
anyway generally supposed to be a 
particle first showing up at much higher 
energies that shall play the role of 
inflaton. So we cannot consider the 
inflaton as supporting our picture
of no new physics, rather on the contrary
the inflaton is rather a problem for our 
picture, but we nevertheless insist that 
it is not truly hopeless with the no new 
physics scenario in spite the need for
an inflaton field. It is namely  not 
totally excluded that it could be the 
Higgs field. Well, really if one could 
solve the anyway for simple reasonable 
model almost unsolvable slow roll 
problem, it would likely be solved also
for the Higgs being the inflaton, and it 
that case one could equally easily use the
Higgs as anyother scalar. That would mean 
that if one first got solved the almost 
universal slow roll problem, the Higgs 
would on an equal footing with any new
invention of a scalar, and thus the 
Standard Model picture up to the Planck 
scale would be o.k..

\section*{Acknowledgement}
It is pleasure to thank Nima Arkani-Hamed 
for informing me about his works
\cite{Nima} about 
Higgs mass prediction from meta-stability 
of the vacuum. Further I want to thank
the collaborators, the work with whom I 
have reviewed, for the inspirations in 
connections with these works, which have 
come 
out in this article, Colin Froggatt,
 - also for telling me about the work 
by Degrassi et al.\cite{Degrassi} - 
Yasutaka 
 Takanishi, Don Bennett, Larisa 
Laperashvili, Masao Ninomiya, Keiichi
Nagao and Roman Nevzorov for discussions
on the mass of our bound state and on 
why MPP.

I also want to thank the Niels Bohr 
Institute for allowing me emeritus status, 
so as to be able to have a working office 
etc. (but during the later part of the 
working time for the present work,
namely after 1st of september 2011,
I have NOT got any salary). I also thank 
Mitja Breskvar for the economical support 
of my very visit and participation in the 
conference of the present proceedings.
Here also thanks for the discussions at 
the workshop.

\end{document}